\documentclass[prl,amsmath,amssymb]{revtex4}
\usepackage{graphicx}

\begin{document}

\title{Black Hole Skyrmions with Negative Cosmological Constant}

\author{Noriko Shiiki}
\email{norikoshiiki@mail.goo.ne.jp}
\author{Nobuyuki Sawado}
\email{sawado@ph.noda.tus.ac.jp}

\affiliation{Department of Physics, Tokyo University of Science, Noda, 
Chiba 278-8510, Japan}

\date{\today}

\begin{abstract}
We study spherically symmetric black hole solutions with 
Skyrme hair in the Einstein-Skyrme theory with a negative 
cosmological constant. 
The dependence of the skyrmion field configuration on the 
cosmological constant is examined. The stability is 
investigated in detail by solving the linearly perturbed 
equation numerically.  
It is shown that there exist linearly stable solutions 
in the branch which represents unstable configuration 
in the asymptotically flat spacetime. 
\end{abstract}

\maketitle 

\section{1. Introduction}
It has been known that the Einstein-Skyrme system possesses 
black hole solutions as well as regular solutions. 
The black hole solution with $B=1$ Skyrme hair was found 
in Refs.~\cite{luckock86,luckock87,droz91,bizon92}. 
It was shown that there exist two fundamental branches 
of the solutions and remarkably one of the branches represents 
stable configuration under linear perturbations~\cite{heusler92,bizon92}. 
This is a counter example to the no-hair conjecture for black holes. 
The no-hair conjecture states that the only allowed degrees of freedom 
of stationary black holes are global charges such as mass, angular momentum 
and electric and magnetic charges~\cite{ruffini71}. 
However, this conjecture concerns only linear field sources and 
asymptotically flat spacetime. 
In fact, if we consider nonlinear field sources such as non-abelian gauge 
fields~\cite{volkov89} and Skyrme fields or asymptotically anti-de Sitter 
spacetime~\cite{torii01}, black hole solutions with hair can exist.   
Following the discovery of black holes $B=1$ Skyrme hair, the $B=2$ Skyrme 
black hole solution with axisymmetry was constructed by us in Ref.~\cite{shiiki04}. 
The review of the black hole solutions with Skyrme hair is given in 
Ref.~\cite{shiiki05}.  
All of these solutions are, however, obtained in the asymptotically flat  
spacetime. 
Thus we consider spherically symmetric black hole solutions with Skyrme 
hair in the asymptotically anti-de Sitter (AdS) spacetime 
with a negative cosmological constant in this paper. 

There has been an increasing interest in the AdS spacetime. 
Especially the AdS black hole is an interesting object from the 
holographic point of view in the form of AdS/CFT 
correspondence~\cite{maldacena98,witten98}.
Maldacena proposed the correspondence between 
the large $N$ limit of conformally invariant $SU(N)$ gauge theory   
and supergravity (string theory) on the product of AdS space with a compact 
manifold in a low energy limit where the dynamics on the brane decouples 
from the bulk. 
Witten elaborated on the idea of the correspondence 
and related the thermodynamics of ${\cal N}=4$ super Yang-Mills 
theory in four dimensions to the thermodynamics of Schwarzschild black holes 
in AdS space~\cite{witten98}. 
The study of the AdS black hole in terms of conformal 
field theory showed that the Hawking-Page phase transition between 
AdS space and a AdS black hole corresponds to the confining-deconfining 
transition in the dual gauge theory~\cite{hawking83}. 
It may be thus useful to consider the Skyrme model which is interpreted 
as QCD in the large $N$ limit in the asymptotically AdS spacetime to understand 
the AdS/CFT correspondence.  
  
Brane world cosmology also indicates that there was a period when  
spacetime was AdS with a negative cosmological constant in the 
early universe (for example, 
see~\cite{randall99,kaloper99,brevik00,karch01}). 
If this scenario is true, AdS black holes should have been produced as 
the primordial black holes. 
The solutions we obtain in this paper provide a semiclassical framework 
to study the interaction of baryons with the AdS primordial black hole. 
 
In the context of the Einstein-Yang-Mills (EYM) theory, 
it was shown that while all black hole solutions are unstable 
for $\Lambda \ge 0$~\cite{zhou90,torii95}, there exist stable black hole 
solutions for $\Lambda < 0$~\cite{winstanley99}. 
We investigate the stability of the AdS Skyrme black hole and 
discuss in detail to see if the cosmological constant 
change the stability of the AdS black hole skyrmion.

\section{2. The Einstein-Skyrme Model}
The Einstein-Skyrme system with the cosmological constant $\Lambda$ 
is defined by the action
\begin{eqnarray}
	{\cal S} = \int d^{4}x \sqrt{-g}\left\{ \frac{1}{16\pi G}(R-2\Lambda) 
	+\frac{F_{\pi}^{2}}{16}g^{\mu\nu}{\rm tr}
	\,(L_{\mu}L_{\nu})+\frac{1}{32e^{2}}g^{\mu\nu}g^{\rho\sigma}{\rm tr}\,
	([L_{\mu},L_{\rho}][L_{\nu},L_{\sigma}])\right\} \label{action} 
\end{eqnarray} 
where $L_{\mu}=U^{\dagger}\partial_{\mu}U$ and $U$ is an $SU(2)$ chiral field. 

We require that the spacetime recovers the AdS solution at infinity 
and parameterize the metric as   
\begin{eqnarray}
	ds^{2}=-e^{2\delta(r)}C(r)dt^{2}+C(r)^{-1}dr^{2}
	+r^{2}(d\theta^{2}+\sin^{2}\theta d\phi^{2})   \label{}
\end{eqnarray}
where  
\begin{eqnarray}
	C(r)=1-\frac{2Gm(r)}{r}-\frac{\Lambda r^{2}}{3} \,. \label{}
\end{eqnarray}  
The topology of AdS spacetime is $S^{1}\times R^{3}$ and hence the 
timelike curves are closed. This can be, however, unwinded if we consider  
the covering spacetime with topology $R^{4}$. 
For the spacetime to recover the anti-de Sitter spacetime asymptotically, 
solutions satisfy 
\begin{eqnarray}
	m(r) \rightarrow {\rm const.} \;\;\;, \;\;\;\;\;
	\delta (r) \rightarrow 0 \;\;\; {\rm as } \;\;\;
	r \rightarrow \infty \,. \label{asym_b}
\end{eqnarray}
At the horizon $r=r_{h}$, we have $C(r_{h})=0$ which reads  
\begin{eqnarray}
	m(r_{h}) = \frac{r_{h}}{2G}-\frac{\Lambda r_{h}^{3}}{6G} 
	\;\;\; , \;\;\;\;\;
	\delta (r_{h})=\delta_{h} \label{}
\end{eqnarray}
where $\delta_{h}$ is provided by hand as a shooting parameter 
so as to satisfy the asymptotic boundary condition in (\ref{asym_b}). 

The $B=1$ skyrmion can be obtained by imposing the hedgehog ansatz on the 
chiral field
\begin{eqnarray}
	U({\vec r})=\cos f(r)+i{\vec n}\cdot{\vec \tau}\sin f(r) \,. \label{}
\end{eqnarray}
Introducing the dimensionless variables  
\begin{eqnarray}
	 x=eF_{\pi}r \; , \;\;\;\;\;\mu(x)=eF_{\pi}G m(r) \;, 
	\;\;\;\;\;{\tilde \Lambda}=\Lambda/e^{2}F_{\pi}^{2}  \label{}
\end{eqnarray} 
with 
\begin{eqnarray}
	C(x)=1-\frac{2\mu(x)}{x}-\frac{{\tilde \Lambda}x^{2}}{3} \,, \label{}
\end{eqnarray}
one gets the energy 
\begin{eqnarray}
	E_{S}&=&4\pi \frac{F_{\pi}}{e} \int \left\{\frac{1}{8}\left(Cf'^{2}
	+\frac{2\sin^{2}f}{x^{2}}\right)+\frac{\sin^{2}f}{2x^{2}}
	\left(2Cf'^{2}+\frac{\sin^{2}f}{x^{2}}\right)\right\}e^{\delta} x^{2} dx \\
	&=& 4\pi \frac{F_{\pi}}{e} \int \left(\frac{1}{8}Cuf'^{2} 
	+\frac{1}{4x^{2}}v\right)e^{\delta} dx  \label{energy}
\end{eqnarray}
where we have defined $u=x^{2}+8\sin^{2}f$ and $v=\sin^{2}f(x^{2}+2\sin^{2}f)$. 

The covariant topological current is defined by  
\begin{eqnarray}
	B^{\mu}=-\frac{\epsilon^{\mu\nu\rho\sigma}}{24\pi^{2}}\frac{1}{\sqrt{-g}}
	{\rm tr}\left(U^{-1}\partial_{\nu}UU^{-1}\partial_{\rho}UU^{-1}
	\partial_{\sigma}U\right) 
	\label{topological_current} \label{baryon_current}
\end{eqnarray}
whose zeroth component corresponds to the baryon number density 
\begin{eqnarray}
	B^{0}=-\frac{1}{2\pi^{2}}e^{-\delta}\frac{f'\sin^{2}f}{r^{2}}\, . \label{}
\end{eqnarray}
We impose the boundary condition on the profile function as  
\begin{eqnarray}
	f(x) \rightarrow 0 \;\;\; {\rm as } \;\;\; x \rightarrow \infty \;, 
	\label{boundary}
\end{eqnarray}
and at the horizon $x=x_{h}$, $f(x_{h})=f_{h}$ is a shooting parameter determined 
so as to satisfy (\ref{boundary}). Then the baryon number becomes    
\begin{eqnarray}
	B=\int \sqrt{-g}\,B^{0} \, d^{3}x = -\frac{2}{\pi}\int_{f_{h}}^{0}
	\sin^{2}f df = \frac{1}{2\pi}(2f_{h}-\sin 2f_{h}) \, . \label{baryon-number}
\end{eqnarray}
If the solution is regular and there is no horizon, the $B=1$ skyrmion 
is recovered with $f_{h}=\pi$. Indeed, as shown later numerically, 
the value of $f_{h}$ approaches to $\pi$ as $x_{h}$ goes to zero.  
However, if the solution is a black hole, $f_{h}$ takes the value 
less than $\pi$, which means that the solution possesses fractional baryonic 
charge. Topologically, the presence of a spherical event horizon alters the 
spacetime topology to be $R^{2}\times S^{2}$ and unwinds the skyrmion. 
Hence the skyrmion absorbed partially by the black hole is topologically 
trivial.  

\begin{figure}
\includegraphics[height=6.5cm, width=8.5cm]{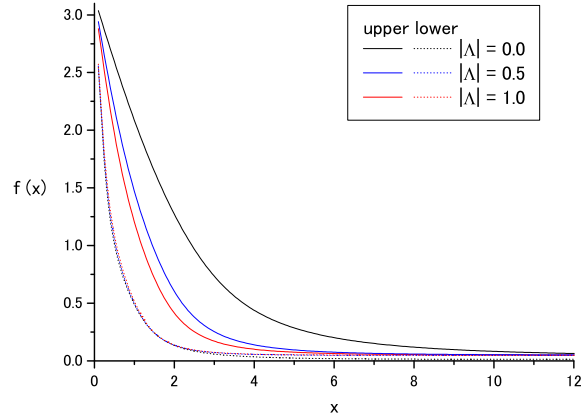}
\caption{\label{fig:F-a002} The profile function $f$ as a 
function of the radial coordinate $x$ for $|{\tilde \Lambda}| 
= 0.0, 0.5, 1.0$ with $x_{h}=0.1$ and $\alpha=0.02$ fixed.}
\end{figure}

\begin{figure}
\includegraphics[height=6.5cm, width=8.5cm]{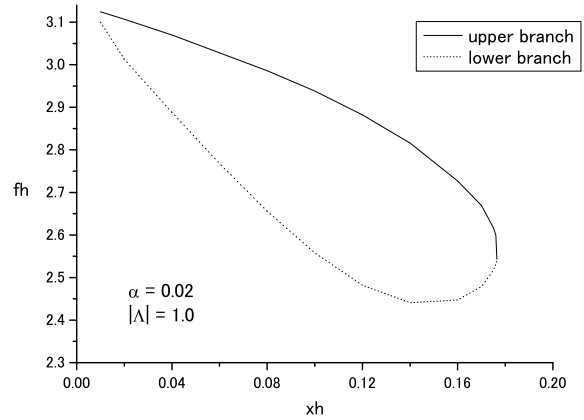}
\caption{\label{fig:xh-fh} The value of the profile at the horizon $f_{h}$ 
as a function of $x_{h}$ with $\alpha =0.02$ and $|{\tilde \Lambda}|= 1.0$ 
fixed.}
\end{figure}

The Einstein equations with a cosmological constant $\Lambda$ takes the form
\begin{eqnarray}
	R_{\mu\nu}-\frac{1}{2}Rg_{\mu\nu}+\Lambda g_{\mu\nu}
	=8\pi GT_{\mu\nu}  \label{}
\end{eqnarray}
which reads     
\begin{eqnarray}
	G_{00}=8\pi GT_{00}-\Lambda g_{00} & \rightarrow &
	1-C-C'x=\frac{\alpha}{4}\left(Cuf'^{2}+\frac{2v}{x^{2}}\right) 
	+{\tilde \Lambda}x^{2} \\
	G_{11}=8\pi GT_{11}-\Lambda g_{11} & \rightarrow &
	-1+C+\frac{(e^{2\delta}C)'}{e^{2\delta}}x=\frac{\alpha}{4}
	\left(Cuf'^{2}-\frac{2v}{x^{2}}\right)-{\tilde \Lambda}x^{2} \,. \label{}
\end{eqnarray}
where we have defined the coupling constant $\alpha=4\pi GF_{\pi}^{2}$. 
Consequently following two equations are obtained for the gravitational fields 
\begin{eqnarray}
       \delta'= \frac{\alpha}{4x}uf'^{2} \label{stab_d} \;,\;\;\;\;\;
	 -(Cx)'+1 = \frac{\alpha}{4}\left(Cuf'^{2}+\frac{2v}{x^{2}}\right)
	 +{\tilde \Lambda}x^{2} \,. 
\end{eqnarray}
The variation of the static energy (\ref{energy}) with respect to the 
profile $f(x)$ leads to the field equation for matter 
\begin{eqnarray}
	f''= \frac{1}{e^{\delta}Cu}\left[-(e^{\delta}Cu)'f'
	+\left(4Cf'^{2}+1+\frac{4\sin^{2}f}{x^{2}}\right)
	e^{\delta}\sin 2f \right] \, .
\end{eqnarray}
Thus the coupled field equations to be solved are given by 
\begin{eqnarray}
	 \delta'&=& \frac{\alpha}{4x}uf'^{2} \label{d-eq}\\ 
	 \mu'&=& \frac{\alpha}{8}\left(Cuf'^{2}+\frac{2v}{x^{2}}\right) 
	 \label{mu-eq} \\
	 f''&=& \frac{1}{e^{\delta}Cu}\left[-(e^{\delta}Cu)'f'
	 +\left(4Cf'^{2}+1+\frac{4\sin^{2}f}{x^{2}}\right)
	 e^{\delta}\sin 2f \right] \, .\label{f-eq}
\end{eqnarray}
In order to determine the boundary conditions on the regular 
event horizon, let us expand the fields around the horizon $x_{h}$ 
\begin{eqnarray}
      \mu &=& \frac{x_{h}}{2}-\frac{{\tilde \Lambda}x_{h}^{3}}{6}
      +\mu_{1}(x-x_{h})+O((x-x_{h})^{2}) \\ 
      f &=& f_{h}+f_{1}(x-x_{h})+O((x-x_{h})^{2}) \\
	\delta &=& \delta_{h}+\delta_{1}(x-x_{h})+O((x-x_{h})^{2}) \,. \label{}
\end{eqnarray}
Inserting them into the field equations~(\ref{d-eq})-(\ref{f-eq}), one obtains
\begin{eqnarray}
      \mu_{1}&=&\frac{\alpha}{4}\left(1+\frac{2\sin^{2}f_{h}}{x_{h}^{2}}
      \right)\sin^{2}f_{h} \\
      f_{1}&=&\frac{x_{h}^{2}+4\sin^{2}f_{h}}{x_{h}(x_{h}^{2}+8\sin^{2}f_{h})
      (1-2\mu_{1}-{\tilde \Lambda}x_{h}^{2})}\sin 2f_{h} \\
	\delta_{1}&=&\frac{\alpha}{4x_{h}}
	(x_{h}^{2}+8\sin^{2}f_{h}) f_{1}^{2}\,. \label{}
\end{eqnarray}
The profile function numerically computed for several values of 
$|{\tilde \Lambda}|$ are shown in Fig.~\ref{fig:F-a002}. The horizon 
radius and the coupling constant are fixed with $x_{h}=0.1$ and $\alpha=0.02$.  
There are two branches of solutions for each value of the cosmological 
constant as well as the coupling constant as was seen in the asymptotically 
flat spacetime case. 
We define the solution with larger (smaller) values of $f_{h}$ 
as upper (lower)-branch. It is shown that the skyrmion shrinks as 
$|{\tilde \Lambda}|$ becomes large in the upper branch. 
In Refs.~\cite{heusler92,bizon92}, one can see that the skyrmion shrinks 
as the coupling constant becomes large in the upper branch. 
Therefore increasing the value of $|{\tilde \Lambda}|$ gives similar effects  
on the skyrmion as increasing the value of the gravitational constant.    
On the other hand, in the lower branch, the change in size is much 
smaller and is almost unrecognizable. In the asymptotically flat spacetime, 
the solution in the lower branch expands in size as the coupling constant 
increases. 
Fig.~\ref{fig:xh-fh} shows the value of $f_{h}$ as a function of $x_{h}$. 
It is observed that $f_{h}$ continuously approaches to $\pi$ as $x_{h}$ goes to 
zero, recovering the $B=1$ regular skyrmion solution at $x_{h}=0$. 

\begin{figure}
\includegraphics[height=6.5cm, width=8.5cm]{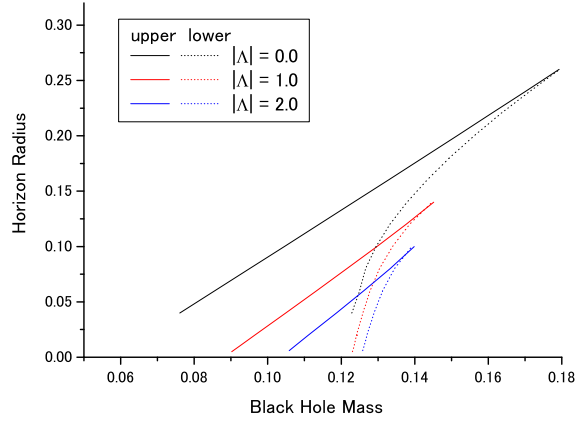}
\caption{\label{fig:Mbh-xh} The horizon radius $x_{h}$ 
as a function of the black hole mass $M_{bh}$ for $|{\tilde \Lambda}|= 
0.0, 1.0, 2.0$.}
\end{figure}

The black hole mass-horizon radius relation is shown in Fig.~\ref{fig:Mbh-xh}. 
It has been shown that the area of an AdS black hole with scalar hair 
proportional to the entropy~\cite{martinez04,hertog05}. 
Thus we infer that this relation still holds for AdS black holes with 
Skyrme hair as  
\begin{eqnarray}
	S=\frac{\pi r_{h}^{2}}{4\hslash G}=\frac{\pi^{2}}{\hslash e^{2}}
	\left(\frac{x_{h}^{2}}{\alpha}\right)\,.
\end{eqnarray} 
One can see that the upper and lower branch correspond to the high- and 
low-entropy branch respectively. The cosmological constant reduces the 
entropy of the black hole. The reduction of the entropy is also seen 
when the coupling constant increases. 

\begin{figure}
\includegraphics[height=6.5cm, width=8.5cm]{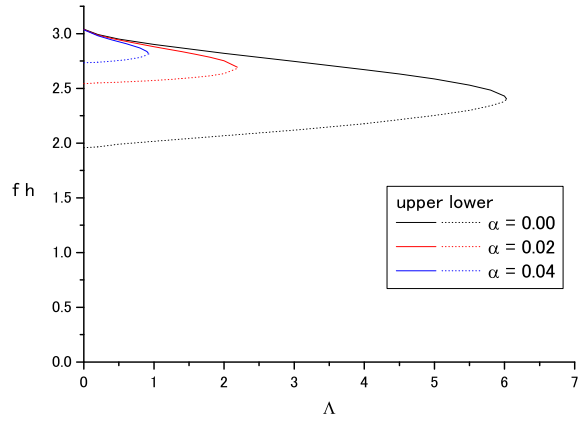}
\caption{\label{fig:L-fh} Parameter $f_{h}$ as a function of $|{\tilde \Lambda}|$ 
for $\alpha = 0.0, 0.02, 0.04$.}
\end{figure}

Fig.~\ref{fig:L-fh} shows the parameter $f_{h}$ as a function of 
$|{\tilde \Lambda}|$ for $\alpha=0.0,\,0.02,\,0.04$ with $x_{h}=0.1$ fixed. 
The value of $f_{h}$ is directly related 
to the baryon number as can be seen from Eq.~(\ref{baryon-number}). 
Thus, in the upper branch, the baryon number becomes smaller  
as $|{\tilde \Lambda}|$ becomes larger, which represents the baryon more absorbed 
by the black hole. In the lower branch, the baryon number slightly 
increases as $|{\tilde \Lambda}|$ becomes large. This result also shows that 
the cosmological constant gives a similar effect on the skyrmion 
as the coupling constant.

\begin{figure}
\includegraphics[height=6.5cm, width=8.5cm]{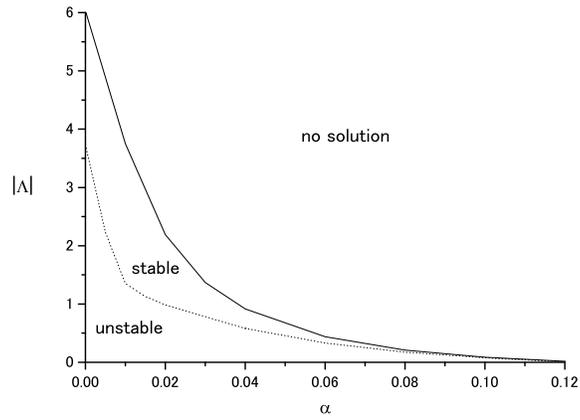}
\caption{\label{fig:a-L} The solid line shows the $\alpha$ dependence of the maximum 
value of $|{\tilde \Lambda}|$ above which there exists no black hole solution. 
The dotted line shows the $\alpha$ dependence of the value of $|{\tilde \Lambda}|$  
above which the lower branch solution changes its stability to become stable.}
\end{figure}

We found the maximum value of $|{\tilde \Lambda}|$ above which there 
exists no solution for each value of the coupling constant.  
In Fig.~\ref{fig:a-L}, the maximum value of $|{\tilde \Lambda}|$ is shown 
as a function of $\alpha$. The maximum value decreases as $\alpha$ increases, 
and at $\alpha=0.126$ it becomes zero. Thus at $\alpha=0.126$, 
the asymptotically AdS solution does not exist and 
only the asymptotically flat solution exists.    

\section{3. Linear Stability Analysis}

In this section we shall examine the linear stability of the black hole 
solutions described in the previous section. 
Let us consider the time-dependent small fluctuation around the 
static classical solutions $f_{0}$, $\delta_{0}$ and $\mu_{0}$ 
\begin{eqnarray}
	f(r,t) &=& f_{0}(r) +f_{1}(r,t) \label{f1} \\
	\delta (r,t)&=& \delta_{0}(r)+\delta_{1}(r,t) \label{d1} \\
	\mu (r,t)&=& \mu_{0}(r)+\mu_{1}(r,t)\,. \label{m1}
\end{eqnarray}
From the time-dependent Einstein-Skyrme action  
\begin{eqnarray}
	{\cal S}=-\frac{\pi e^{2}F_{\pi}^{4}}{2}\int \left[(-\frac{1}{e^{\delta}C}
	{\dot f}^{2}+Cf'^{2})u + v \right]e^{\delta} dx \, , \label{time-action}
\end{eqnarray}
one obtains the time-dependent field equation as  
\begin{eqnarray}
	(e^{\delta}Cuf')'+\frac{1}{2}\left(\frac{1}{e^{\delta}C}{\dot f}^{2}
	-e^{\delta}Cf'^{2}\right)u_{f}-\frac{e^{\delta}v_{f}}{x^{2}}
	=\frac{1}{e^{\delta}C}u{\ddot f} \label{time-field-eq}
\end{eqnarray}
where we have defined $u_{f}=\delta u/\delta f$ and $v_{f}=\delta v/\delta f$. 

\begin{figure}
\includegraphics[height=6.5cm, width=8.5cm]{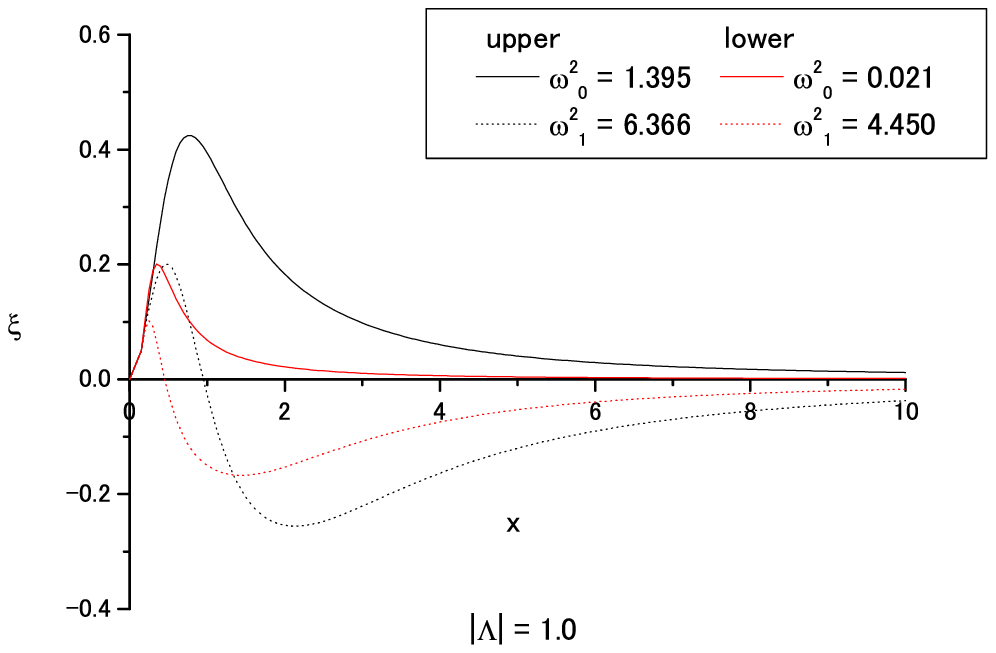}
\caption{\label{fig:wave(L=1)}  Wave function $\xi$ for the upper 
and lower branch with $|{\tilde \Lambda}|=1$ and $\alpha=0.02$. }
\end{figure}

\begin{figure}
\includegraphics[height=6.5cm, width=8.5cm]{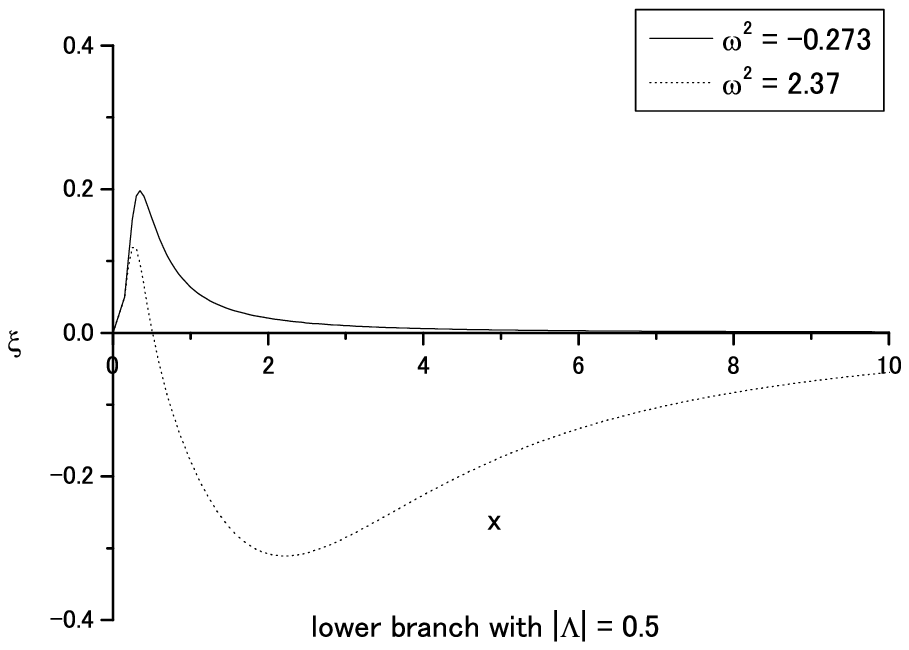}
\caption{\label{fig:wave(L=05)}  Wave function $\xi$ for the lower branch with 
$|{\tilde \Lambda}|=0.5$ and $\alpha=0.02$. }
\end{figure}

\begin{figure}
\includegraphics[height=6.5cm, width=8.5cm]{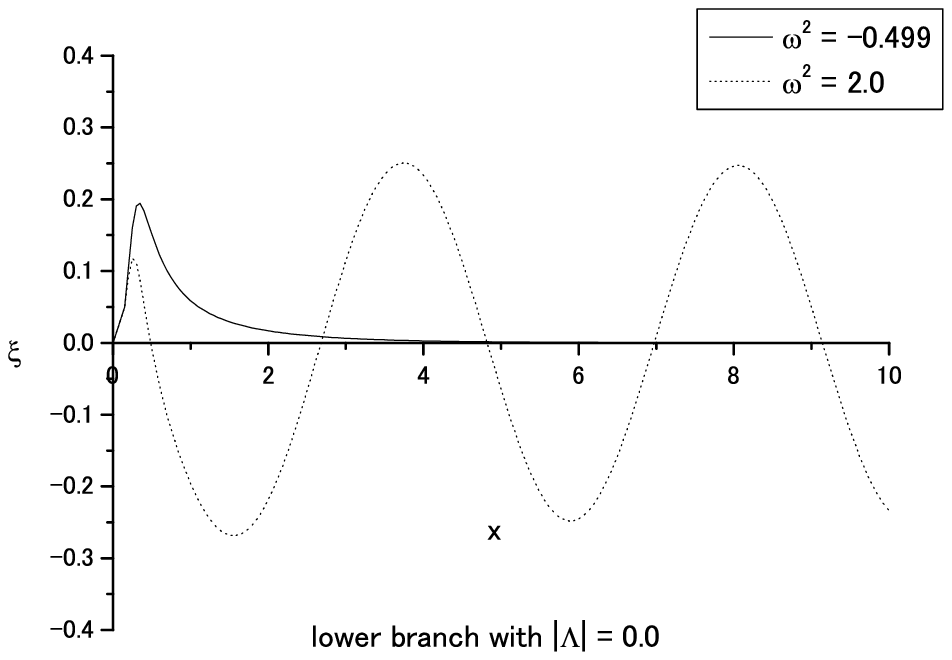}
\caption{\label{fig:wave-low-L0}  Wave function $\xi$ for the lower branch with 
$|{\tilde \Lambda}|=0.0$ and $\alpha=0.02$. }
\end{figure}

The time-dependent Einstein equations are derived as 
\begin{eqnarray}
	G_{00}=8\pi GT_{00} & \rightarrow &
	1-C-C'x=\frac{\alpha}{4}\left[\left(\frac{1}{e^{2\delta}C}{\dot f}^{2}
	+Cf'^{2}\right)u+\frac{2v}{x^{2}}\right]\\
	G_{11}=8\pi GT_{11} & \rightarrow &
	-1+C+\frac{(e^{2\delta}C)'}{e^{2\delta}}x=\frac{\alpha}{4}
	\left[\left(\frac{1}{e^{2\delta}C}{\dot f}^{2}+Cf'^{2}\right)u
	-\frac{2v}{x^{2}}\right] \label{}
\end{eqnarray}
which reads the following two equations  
\begin{eqnarray}
	&& \delta'= \frac{\alpha}{4x}\left(\frac{1}{e^{2\delta}C^{2}}{\dot f}^{2}
	+f'^{2}\right)u \label{stab_d} \\ 
	&& -(Cx)'+1 = \frac{\alpha}{2x^{2}}v+C\delta'x \label{stab_c} \,.\label{}
\end{eqnarray}
Substituting Eqs.~(\ref{f1})-(\ref{m1}) into Eqs.~(\ref{stab_d}) and (\ref{stab_c}) 
gives the linearized equations   
\begin{eqnarray}
        \delta_{1}' &=& \frac{\alpha}{2x}(2u_{0}f_{0}'f_{1}'
       +u_{f_{0}}f_{0}'^{2}f_{1}) \label{ddel1} \\
	  -(e^{\delta_{0}}C_{1}x)'&=& \frac{\alpha}{2x^{2}}e^{\delta_{0}}
	v_{f_{0}}f_{1}+e^{\delta_{0}}C_{0}\delta_{1}'x \, .\label{dc1}
\end{eqnarray}
Eq.~(\ref{ddel1}) and the classical field equation derived 
from Eq.~(\ref{time-field-eq}) which is 
\begin{eqnarray}
	 \frac{e^{\delta_{0}}v_{f_{0}}}{x^{2}}=(e^{\delta_{0}}C_{0}u_{0}f_{0}')'
	-\frac{1}{2}e^{\delta_{0}}C_{0}u_{f_{0}}f_{0}'^{2}\, , \label{static-field}
\end{eqnarray}
are inserted into Eq.~(\ref{dc1}) and resultantly one gets  
\begin{eqnarray}
	-(e^{\delta_{0}}C_{1}x)'=\frac{\alpha}{2}(e^{\delta_{0}}
	C_{0}u_{0}f_{0}'f_{1})'  \label{}
\end{eqnarray}
which can be integrated immediately to obtain 
\begin{eqnarray}
	C_{1}=-\frac{\alpha}{2x}C_{0}u_{0}f_{0}'f_{1} \,. \label{c1}
\end{eqnarray}
Similarly let us linearize the field equation~(\ref{time-field-eq}). 
Using Eqs.~(\ref{ddel1}), (\ref{static-field}) and (\ref{c1}), one arrives at   
\begin{eqnarray}
	(e^{\delta_{0}}C_{0}u_{0}f_{1}')'-U_{0}f_{1}=\frac{1}{e^{\delta_{0}}C_{0}}
	u_{0}{\ddot f_{1}} \label{f1-eq}
\end{eqnarray}
where 
\begin{eqnarray}
	U_{0}&=&-(e^{\delta_{0}}C_{0}u_{f_{0}}f_{0}')'
	+\left(\frac{\alpha}{2x}e^{\delta_{0}}C_{0}u_{0}^{2}
	f_{0}'^{2}\right)'-\frac{\alpha}{2x}e^{\delta_{0}}C_{0}
	u_{0}u_{f_{0}}f_{0}'^{3} \nonumber \\
	&& +\frac{1}{2}e^{\delta_{0}}C_{0}u_{ff_{0}}f_{0}'^{2}
	+\frac{e^{\delta_{0}}v_{ff_{0}}}{x^{2}}\,. \label{}
\end{eqnarray}
Setting $f_{1}=\xi (x)e^{i\omega t}/\sqrt{u_{0}}$ , we derive from Eq.~(\ref{f1-eq}) as  
\begin{eqnarray}
	-(e^{\delta_{0}}C_{0}\xi')'+\left[\frac{1}{2\sqrt{u_{0}}}
	\left(e^{\delta_{0}}C_{0}\frac{u_{0}'}{\sqrt{u_{0}}}\right)'
	+\frac{1}{u_{0}}U_{0}\right] \xi =\omega^{2}
	\frac{1}{e^{\delta_{0}}C_{0}}\xi \,. \label{eigen-eq}
\end{eqnarray}
Let us introduce the tortoise coordinate $x^{*}$ such that 
\begin{eqnarray}
	\frac{dx^{*}}{dx}=\frac{1}{e^{\delta_{0}}C_{0}} \label{}
\end{eqnarray}
with $-\infty < x^{*} < +\infty$. Eq.~(\ref{eigen-eq}) is then reduced 
to the Strum-Liouville equation 
\begin{eqnarray}
	-\frac{d^{2}\xi}{dx^{*2}}+{\hat U}_{0}\xi =\omega^{2}\xi  \label{}
\end{eqnarray}
where
\begin{eqnarray}
	{\hat U}_{0}=e^{\delta_{0}}C_{0}\left[\frac{1}{2\sqrt{u_{0}}}
	\left(e^{\delta_{0}}C_{0}\frac{u_{0}'}{\sqrt{u_{0}}}\right)'
	+\frac{1}{u_{0}}U_{0} \right]\,. \label{strum}
\end{eqnarray}
The solution is linearly stable if there exist no negative eigenvalues 
since imaginary $\omega$ represent exponentially growing modes. 
Unfortunately the potential ${\hat U}_{0}$ has a complicated 
form and we are unable to discuss the stability analytically.  
Thus we solve the wave equation (\ref{eigen-eq}) numerically under the 
boundary conditions $\xi \rightarrow 0$ as $x \rightarrow x_{h}$ and 
$x \rightarrow \infty$ which ensure the norm of the wave function to be finite.  
We show the wave function in Fig.~\ref{fig:wave(L=1)} for $|{\tilde \Lambda}|=1$ 
with $\alpha =0.02$ and $x_{h}=1$ fixed. The ground state corresponds   
to the wave function with no node. The first excited state corresponds to 
the wave function with one node. 
Remarkably both the lower and upper branch solution have ground states  
with $\omega^{2}$ positive and hence they are stable. 
Fig.~\ref{fig:wave(L=05)} shows the wave function of the lower branch 
for $|{\tilde \Lambda}|=0.5$. The ground state has a negative eigenvalue 
$(\omega^{2}=-0.273)$ and thus the solution is unstable. 
We have found the critical value of $|{\tilde \Lambda}|$ at which the 
stability of the lower branch solution changes and shown in Fig.~\ref{fig:a-L} 
as a dotted line.  
For $|{\tilde \Lambda}|=0$, the lower branch solution has one negative mode 
and for $\omega^{2}\ge 0$, the mode becomes continuous 
(see Fig.~\ref{fig:wave-low-L0}). Hence it is unstable for all values of 
the parameters $\alpha$ and $x_{h}$, which confirms the previous results 
in Ref.~\cite{bizon92,heusler92}.  

The results we have obtained indicate that the cosmological constant 
stabilizes the black hole skyrmion. 
This is consistent with the result of the EYM black hole where it was 
shown that the EYM black hole is unstable for 
$\Lambda \ge 0$~\cite{zhou90,torii95}, but 
the stable EYM black hole exists for $\Lambda <0$~\cite{winstanley99}. 

Another important feature for the black hole skyrmion with $\Lambda <0$ 
is that only discrete modes exist in both of the branches. 
It is because the potential behaves asymptotically as $x^{2}$, meaning 
all the eigenvalues are discretized analogous to the harmonic oscillator 
eigenvalues. 
 
\section{4. Conclusions}

We have studied the black hole skyrmion with a negative cosmological 
constant. There exists two fundamental branches of the solutions; the 
upper- and lower-branch corresponding to the high- and low-entropy branch 
respectively. 
The increase in the absolute value of the cosmological constant 
$|\Lambda|$ gives similar effects on the skyrmion as increasing effectively 
the value of the gravitational constant fixing the pion decay constant 
to be the experimental value.  
The skyrmion shrinks and the baryon number is more absorbed by the black hole 
as $|\Lambda|$ increases.  
There is the maximum value of $|\Lambda|$ above which no solution 
exists for each value of the coupling constant. 
In particular, we observed that at $\alpha = 0.126$, no asymptotically AdS solution 
exists and only the asymptotically flat solution exists.   
 
The linear stability was examined in detail by solving the linear perturbed 
wave equation numerically. 
In the asymptotically flat case, the upper branch is stable and the lower 
branch is unstable. However, in the AdS case, surprisingly there exist stable 
solutions even in the lower branch depending on the value of $|\Lambda|$. 
The observation that the negative cosmological constant stabilizes the black hole 
skyrmion solution is consistent with the result for the EYM black hole where 
there exist stable solutions only when $\Lambda <0$~\cite{zhou90,torii95,
winstanley99}.   
This implies that the catastrophe theory for the stability analysis of hairy 
black holes~\cite{maeda94} may not be applicable to the black hole solution 
with a negative cosmological constant.   

The solutions we obtained in this paper provide a semiclassical framework 
to study the interaction of baryons with the AdS primordial black hole. 
If the AdS primordial black holes were created in the early universe 
as indicated in Refs.~\cite{randall99,kaloper99,brevik00,karch01}, 
it would have induced baryon decay. The Einstein-Skyrme theory should 
be useful as a simple framework to study such decay process. 
 
Since the Skyrme model corresponds to QCD in the large $N$ limit, 
it may be worth understanding the Skyrme model in the 
context of the string theory. With the assumption that the brane is AdS, 
the extension of recently discovered brane-skyrmions to AdS brane-skyrmions 
would be one of the possibilities~\cite{cembranos01}.



\end{document}